\documentclass[aps,prl,twocolumn,superscriptaddress,floatfix,longbibliography,showpacs]{revtex4-1}

\usepackage{graphicx}
\usepackage{dcolumn}
\usepackage{bm}
\usepackage{amssymb}
\usepackage{microtype}
\usepackage{gensymb}
\usepackage{xcolor}

\begin{document}

\title{Effect of synthesis conditions on the electrical resistivity of TiSe$_2$}

\author{Jaime M. Moya}
\email[Email: ]{jmm25@rice.edu}
\affiliation{Applied Physics Program, Rice University}
\affiliation{Department of Physics and Astronomy, Rice University}

\author{C.-L. Huang}
\affiliation{Department of Physics and Astronomy, Rice University}

\author{Jesse Choe}
\affiliation{Department of Electrical and Computer Engineering, Rice University}

\author{Gelu Costin}
\affiliation{Department of Earth, Environmental and Planetary Sciences, Rice University}

\author{Matthew S. Foster}
\affiliation{Department of Physics and Astronomy, Rice University}

\author{E. Morosan}
\affiliation{Department of Physics and Astronomy, Rice University}

\date{\today}

\begin{abstract}

Dilute impurities and growth conditions can drastically affect the transport properties of TiSe$_2$, especially below the charge density wave transition. In this paper, we discuss the effects of cooling rate, annealing time and annealing temperature on the transport properties of TiSe$_2$: slow cooling of polycrystalline TiSe$_2$ post-synthesis drastically increases the low temperature resistivity, which is in contrast to the metallic behavior of single-crystalline TiSe$_2$ due to charge doping from the residual iodine transport agent. A logarithmic increase of resistivity upon cooling and negative magnetoresistance with a sharp cusp around zero field are observed for the first time for the polycrystalline TiSe$_2$ samples, pointing to weak-localization effects due to low dimensionality. Annealing at low temperatures has a similar, but less drastic effect. Furthermore, rapid quenching of the polycrystalline samples from high temperatures freezes in disorder, leading to a decrease in the low temperature resistivity. 

\end{abstract}

\maketitle

\section{\label{sec:level1}INTRODUCTION}

Transition metal dichalcogenides (TMDCs) are a class of layered quasi-two dimensional materials.  Owing to their low dimensionality, they span a vast area of physical properties.  TiSe$_2$ is one such TMDC that has attracted lots of attention due to its complex electronic properties, including charge ordering \cite{Salvo1976}, superconductivity with intercalation of copper or palladium \cite{Morosan2006, Morosan2010}, and with the application of pressure \cite{Kusmartseva2009} or electrostatic gating \cite{Li2016}. On the other extreme, TiSe$_2$ becomes insulating with platinum doping \cite{Chen2015a}, and displays potential Luttinger liquid states within domain boundaries \cite{Choe2019} revealing the versatility of the chemical tuning of this TMDC compound.

The origin of the charge density wave (CDW) transition, occurring in TiSe$_2$ around 202 K \cite{Salvo1976}, has been an ongoing debate for decades, with proposed mechanisms including an excitonic insulator phase \cite{Jerome1967} and the band-type Jahn-Teller effect \cite{Hughes1977}. For the former, it can arise either in a small band gap semiconductor or a semimetal \cite{Jerome1967}. Below the CDW transition, angle-resolved photoemission spectroscopy (ARPES) experiments point to a small bandgap. However, the normal state indirect band gap is small, and its absolute value (positive or negative) is still under debate \cite{Watson2018, Bachrach1976, Traum1978, Pillo2000, Kidd2002, Rossnagel2002, Cercellier2007, May2011,  Chen2015}. The latter proposed CDW mechanism is independent of the free carrier concentration \cite{Hughes1977}, and this cannot account for the incommensurate diffraction spots seen in TiSe$_2$ \cite{Salvo1976}.  Recent experimental evidence favors the excitonic insulator scenario \cite{Hildebrand2016, Monney2009, Monney2011, Cercellier2007, Cazzaniga2012, Kogar2017}, but theories predict that the exciton condensation can either be a superfluid \cite{Snoke2002}, or an insulator \cite{Kohn1970}. Most recently, Watson $et$ $al.$ presented resistivity simulations, assuming a semiconducting normal state \cite{Watson2019}. Even without implementing CDW physics, these simulations reproduced the anomalous peak observed in experiments around 150~K. Huang \textit{et al}. \cite{Huang2017} showed insulating behavior for their polycrystalline TiSe$_2$ samples closest to stoichiometry, with metalicity induced by increasing Se deficiency \cite{Huang2017}, while Campbell \textit{et al}. recently revealed insulating behavior in iodine-free single crystals \cite{Campbell2018}. Historically though, single crystal samples grown by iodine vapor transport have shown metallic behavior in resistivity \cite{Salvo1976, Rossnagel2002, Li2007}. Bearing all of the above in mind, it is essential to reach experimental resolution of the intrinsic ground state of TiSe$_2$.

\begin{figure}[t!]
  \includegraphics[width=\linewidth]{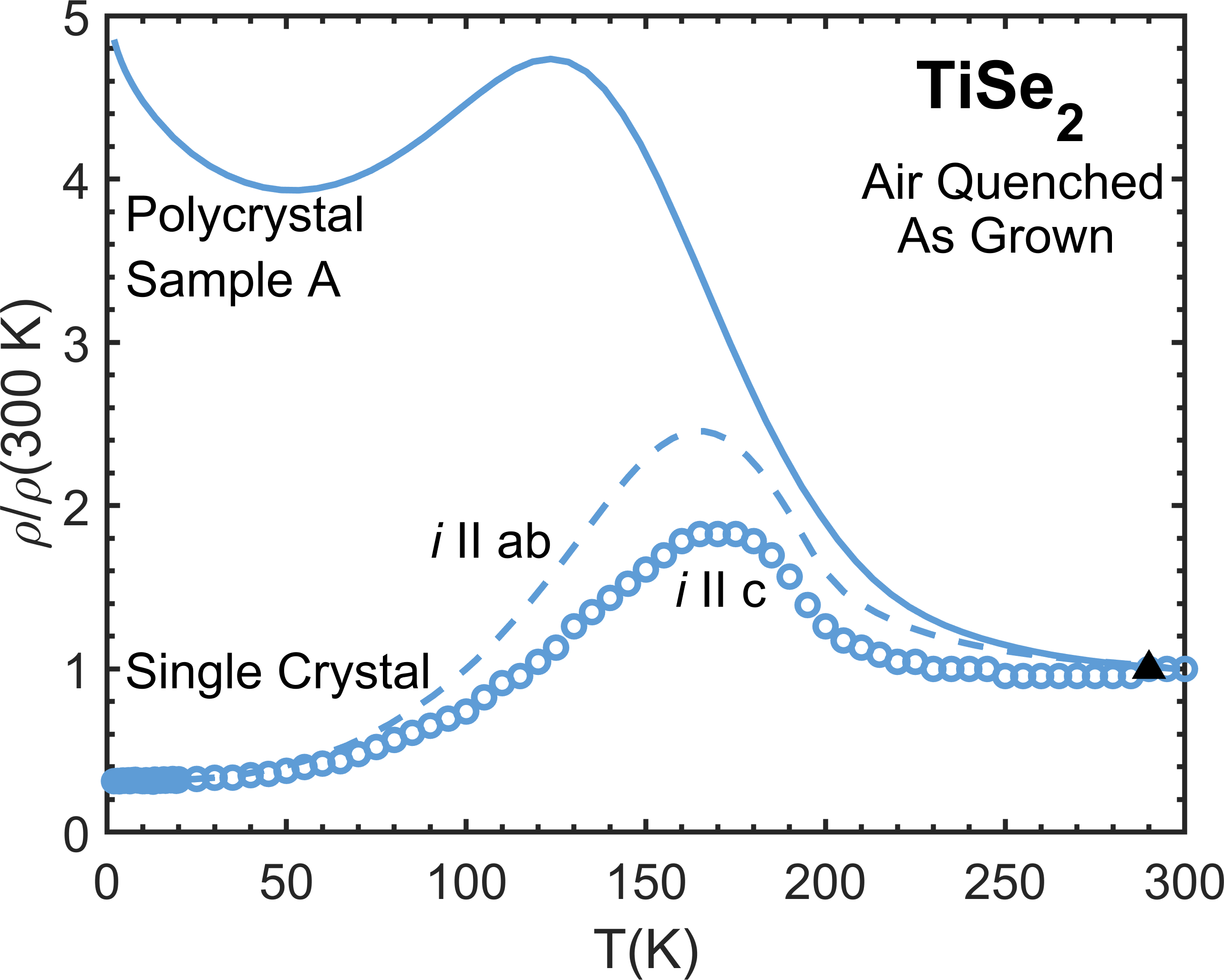}
  \caption{A comparison of the resistivity (normalized to room temperature values) for iodine-grown TiSe$_2$ single crystals with the current $i~\parallel~ab$ (dashed line) or $i~\parallel~c$ (open circles), and polycrystalline (solid line). The full triangle is used to identify samples that are `As Grown, Air Quenched' throughout the text.} 
 
  \label{fig:singlepolycomp}
\end{figure}

One problem faced in studying TiSe$_2$ is the inconsistency in the physical properties from sample to sample. The temperature-dependent resistivity $\rho$(T) shows discrepancy between single-crystalline TiSe$_2$ grown by I$_2$ vapor transport  \cite{Salvo1976, Rossnagel2002, Li2007} and polycrystalline TiSe$_2$, synthesized by solid state reaction \cite{Chen2015, Morosan2006, Huang2017}.  This is illustrated by the normalized $\rho$(T) data of TiSe$_2$ in Fig.~\ref{fig:singlepolycomp}. Even though the $\rho$(T) behavior is qualitatively similar between single-crystalline and polycrystalline samples with a local maximum between 100 and 200~K, at the lower temperatures $\rho$(T) varies drastically: metallic behavior (d$\rho$/dT $> 0$) is registered in the single-crystalline sample (dashed line and open circles), explained by either a doped semiconductor picture \cite{Watson2019} or partial gapping of the Fermi surface \cite{Salvo1976}, while semiconductor-like behavior (d$\rho$/dT $< 0$) is found in the polycrystalline sample (solid line). To our knowledge, no systematic study of this discrepancy exists. It is imperative to understand the intrinsic properties of TiSe$_2$, and the effect of the synthesis conditions on the observed resistivity measurements, before the more complex effects of chemical doping, intercalation, or pressure can be understood.

It is well known that, for TiSe$_2$ single crystals, the transport agent iodine might partially substitute for Se and dope the system \cite{Salvo1976,Rossnagel2002}. Se deficiency also serves as a method of self-doping \cite{Chen2016}. Both dopants presumably contribute additional density of states near the Fermi surface and hence enhance the conductivity on cooling. Here, we report systematic variations in the electrical transport properties of polycrystalline TiSe$_2$ (without doping or Se deficiency), as a function of \textit{cooling rate, annealing time}, and \textit{annealing temperature}. By decreasing the rate at which samples are cooled post-synthesis, an increase in low temperature resistivity is observed. We surmise that the observed logarithmic temperature dependence is due to weak-localization (WL) effects in low dimensional systems. Annealing polycrystalline samples post-synthesis at low temperatures (200$^\circ$C) has a similar, but less drastic effect. Our results are consistent with a possible intrinsic semiconducting ground state in TiSe$_2$.

\section{\label{sec:level1} METHODS}
Polycrystalline samples of TiSe$_2$ were synthesized by solid state reaction with a Ti:Se ratio of 1:2.02. The excess Se was added to compensate for the partial evaporation inherent during synthesis. The samples were sealed in quartz ampoules under partial Argon atmosphere and heated at 50$^\circ$C/hr to 650$^\circ$C, followed by a 48 hour dwell at this temperature. Subsequently, the samples were either cooled at different rates, or annealed at different temperatures or different times under partial Argon atmosphere. TiSe$_2$ single crystals were grown by chemical vapor transport with I$_2$ as the transport agent. Ground elemental Ti and Se were sealed in quartz tubes with a ratio of 1:2.02 and 5 mg/cm$^3$ of iodine. The tubes were then placed in a 550$^\circ$C - 650$^\circ$C temperature gradient and held for 14 days, followed by controlled cooling to room temperature.

Structural characterization was done using a Bruker X-ray diffractometer with Cu k$_\alpha$ radiation. Refinements were performed using the FullProf software package \cite{Carvajal1990}. The quantitative chemical composition was
determined by electron probe microanalysis (EPMA) using a
JEOL JXA 8530F Hyperprobe located at Rice University,
Department of Earth, Environmental and Planetary Sciences,
and equipped with a Schottky field emitter and five
wavelength dispersive spectrometers. The analytical
conditions were set to 15 kV accelerating voltage, 20 nA beam
current, and beam spot size ($\sim$300 nm). The Se L$_\alpha$ and Ti K$_\alpha$
X-ray lines were simultaneously measured using counting
times of 10 seconds per peak and 5 seconds per each lower
and upper background, respectively. Each element was
simultaneously measured on two different spectrometers in
order to increase the accuracy and the statistics of the
measurement. Se L$_\alpha$ was analyzed on two TAP diffracting
crystals, and Ti K$_\alpha$ was analyzed on a PETL and a LiFH
diffracting crystal, respectively. The standards used to
calibrate the spectrometers were Se metal (Se = 99.9990 wt.
$\%$) and rutile (TiO$_2$, where Ti = 59.9400 wt. $\%$). For
quantification, the ZAF matrix correction was used. The error
of analysis, determined after analyzing secondary standards is
below 2$\%$. The instrumental standard deviation (1$\sigma$) for Se
and Ti in each analysis is 0.24$\%$ and 0.47$\%$, respectively. The
quantitative analyses given in element wt. $\%$ were
converted to atomic ratios, and then the stoichiometry of
the analyzed compound was normalized to one Ti atom.

Polycrystalline samples were pressed into pellets without sintering, and shaped into bars for resistivity measurements. DC electrical resistivity measurements were made in a Quantum Design Physical Properties Measurement System with a standard four-point probe technique for temperatures $2-300$ K. The technique described in Ref.~\onlinecite{Maeno1994} was used for resistance measurements with curent $i\parallel~c$. Hall coefficient measurements were performed at constant temperature for selected temperatures sweeping fields from -9~T to 9~T to extract the Hall resistance.
\section{\label{sec:level1} Results and Discussion}
\subsection{\label{sec:level2} Post synthesis cooling rate \textit{r}}
When trying to improve the quality of crystals (\textit{e.g.} decrease extrinsic disorder), two commonly used techniques for metals are: (i) slow cooling to avoid quenching in disorder, and (ii) post synthesis annealing below the synthesis temperature to relieve microstrain and increase grain size \cite{Cullity}. In the present study, both methods were employed to minimize disorder. By contrast, quenching from high temperature was used to study the effect of enhanced disorder.

\begin{figure}[b!]
  \includegraphics[width=\linewidth]{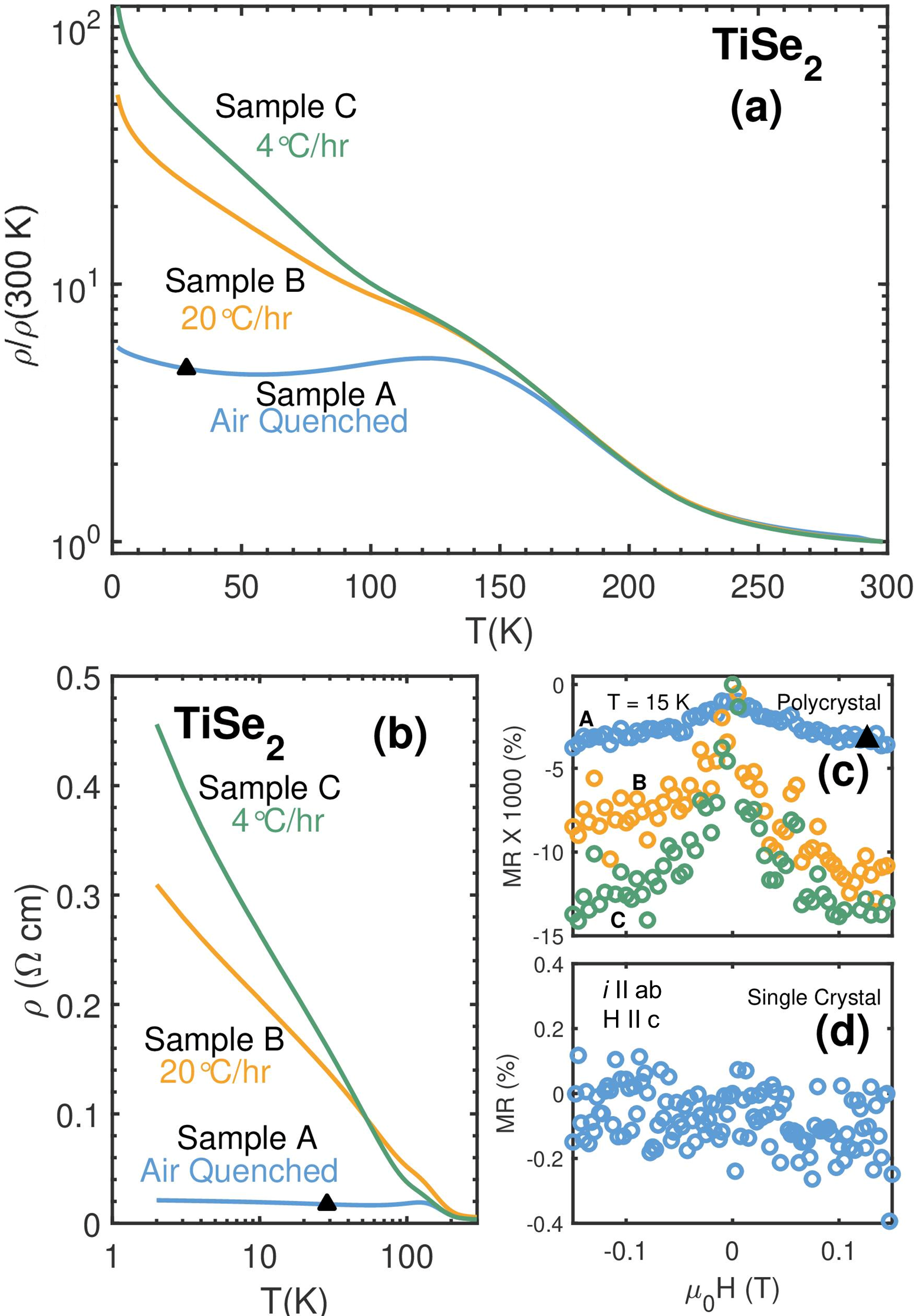}
  \caption{(a) A comparison of the normalized resistivity $\rho/\rho$(300 K) as a function of temperature for samples: A, air quenched; B, cooled at 20$^\circ$C/hr; C, cooled at 4$^\circ$C/hr. (b) A semi-logarithmic plot of $\rho$(T). (c) and (d) show the magnetoresistance MR measured at 15~K for the polycrystalline samples A-C and the single crystal, respectively.}
  \label{fig:SlowCoolResistivity}
\end{figure}

\begin{figure}[h!]
  \includegraphics[width=\linewidth]{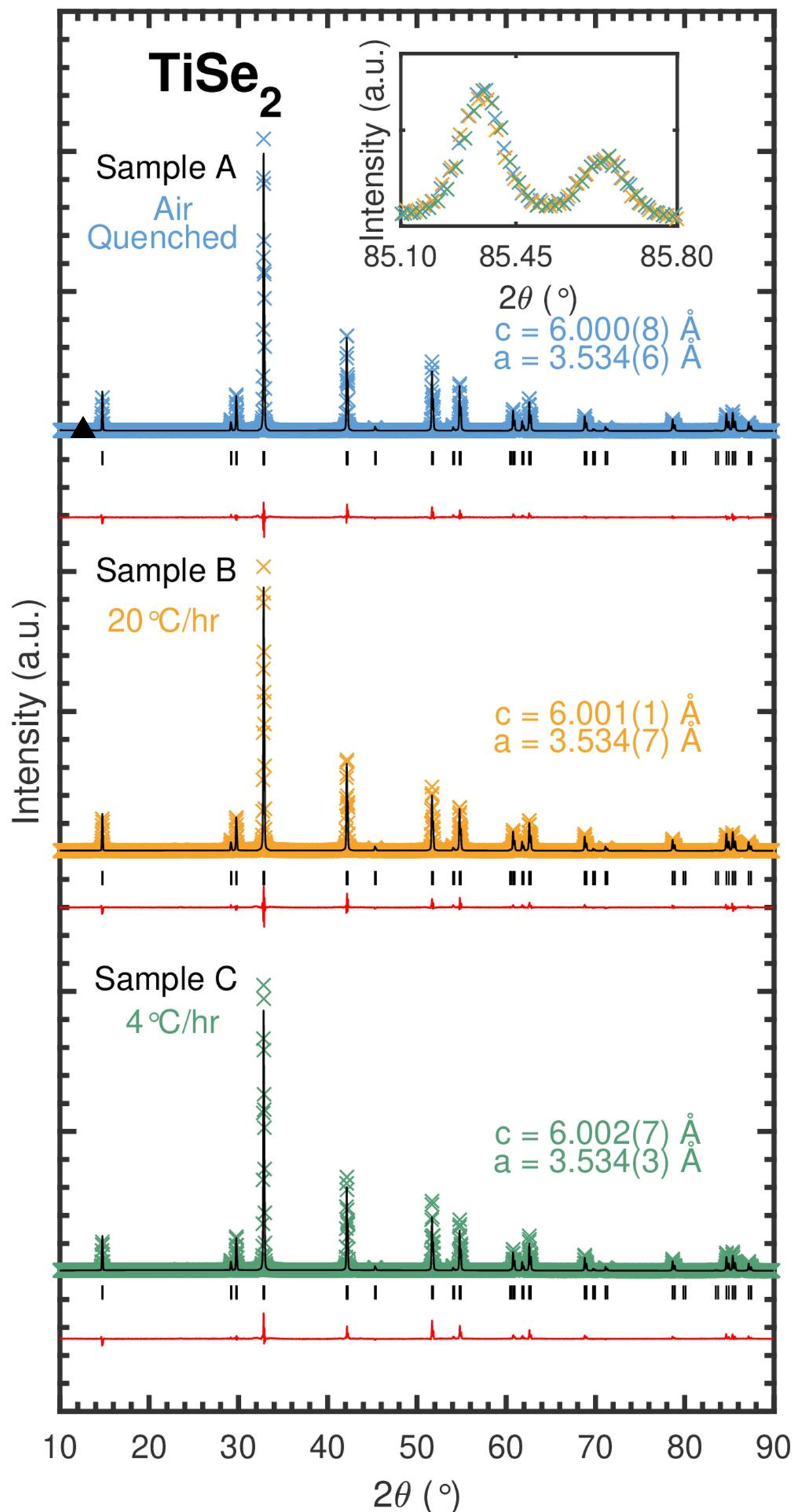}
  \caption{Room temperature powder X-ray diffraction patterns for the polycrystalline samples (symbols), with refinements shown as solid black lines, and the difference between measurement and calculation shown in red.  The vertical marks below each pattern correspond to the calculated peak position for TiSe$_2$. Inset: a zoomed in view of all three of the measured patterns plotted. }
  \label{fig:SlowCoolxray}
\end{figure}

The first experiment was dedicated to testing the effect of the cooling rate \textit{r} post synthesis on the electrical resistivity. Three samples were synthesized as described in the Methods. Sample A was air quenched ($r~>~2000^\circ$C/hr), sample B was fast-cooled to room temperature at a rate $r~=$ 20$^\circ$C/hr, and sample C was slow-cooled at $r~=$ 4$^\circ$C/hr. The scaled semi-log $\rho(T)/\rho$(300 K) plot is displayed in Fig. \ref{fig:SlowCoolResistivity}(a). While all three samples show a nearly 5 time increase in $\rho/\rho$(300 K) on cooling to 150~K, the air-quenched sample A displays a broad local minimum centered around 60~K, while both samples B and C exhibit nearly two orders of magnitude resistivity increase down to 2 K. Hall coefficient values (not shown) are negative at low temperatures, consistent with reported data\cite{Salvo1976,Campbell2018}. This rules out the possibility of a change in dominant carrier type as the cause of change in the low temperature resistivity. The large change in the resistivity as a function of cooling rate prompted the need to check sample composition for possible non-stoichiometry. The results of the EPMA analysis, displayed in Table \ref{Tab:EMPA1}, indicate that all three samples are stoichiometric (to within a 1\% error). This does not rule out that the resistivity changes between the three samples may be due to composition variations below the EPMA resolution limit, or, as discussed below, conductive grain boundaries and WL effects. Room temperature X-ray diffraction data (Fig. \ref{fig:SlowCoolxray}) does not show any measurable change in either the peak position or peak shape among the three samples, consistent with invariable lattice parameters.

\begin{center}
      \begin{table}[htbp]
          \begin{tabular}{l || c c }
              \multicolumn{2}{c}{} \\ \hline
              Cooling rate $r$ ($^\circ$C/hr) & Se   \\ \hline
              {A: $>$ 2000 (air quench)} & ${2.02\pm0.01}$   \\
              {B: 20} & ${2.01\pm0.01}$ \\
              {C: 4}  & ${2.00\pm0.01}$ \\
              \hline \hline
          \end{tabular}
          \caption{EPMA results for polycrystalline TiSe$_2$ with variable cooling rates post synthesis corresponding to Figs. \ref{fig:SlowCoolResistivity}  and \ref{fig:SlowCoolxray}. Data is normalized to 1 Ti.} \label{Tab:EMPA1}
      \end{table}
  \end{center}

When plotting $\rho(T)$ on a semi-log scale (Fig.~\ref{fig:SlowCoolResistivity}(b)), all three samples A-C show a $-$lnT dependence of $\rho(T)$ upon cooling below the broad local maximum near 150~K. Since no magnetic impurities are present in any of the samples, the $-$lnT increase of $\rho$ cannot be attributed to Kondo or other extrinsic magnetic effects. In TiSe$_2$, the low dimensionality enhances two quantum corrections to the resistivity: Altshuler-Aronov corrections due to the coherent scattering of electrons by impurity-induced Friedel oscillations \cite{AA1985,Aleiner2001,Lara2011}, and WL due to self-intersecting scattering paths \cite{Hikami1980,Lee1985}. Upon an application of finite transverse magnetic field $H~\perp~i$, the shape of the magnetoresistance MR $=~ [\rho(H)-\rho(0)]/\rho(0)$ is insensitive to Altshuler-Aronov corrections, while WL can be suppressed in finite magnetic fields leading to a negative MR. Fig.~\ref{fig:SlowCoolResistivity}(c) shows a pronounced peak of MR centered at zero field for samples A-C, which is typical for WL effects. However, the absolute MR values for the different samples reflect not only the WL effects, but also extrinsic effects likely due to the different cooling rates. Therefore, possible explanations for the low temperature increase in resistivity $\rho(T)$ with decreasing $r$ include disorder, or grain boundaries more conductive than TiSe$_2$. It has been shown that grain boundaries in polycrystalline samples can be conductive \cite{visoly2006understanding}. Slow cooling (small \textit{r}) would be expected to increase grain size, reducing disorder and the number of grain boundaries, and thus increasing the low temperature resistivity.

For comparison, the single crystal sample with iodine inclusions does not show WL behavior either in $\rho$(T) or in MR (Fig.~\ref{fig:SlowCoolResistivity}(d)). EPMA reveals a 1$\%$ iodine impurity per formula unit in the single crysalline samples.
In our single crystal sample, the iodine inclusions might dope the system and dominate the transport property which leads to a suppression of WL behavior. A recent electrical transport study on iodine-free TiSe$_2$ single crystals 
does show a large increase in electrical resistivity on cooling, qualitatively consistent with what is seen in our polycrystalline Samples B and C \cite{Campbell2018}. It will be informative to investigate the magnetic field effects on the transport properties in these iodine-free single crystals to quantitatively analyze the characteristic parameters from the WL correction. The WL effect noted here for the first time in TiSe$_2$ had been previously reported in another TMDC, VSe$_2$\cite{Cao2017}.

Cooling samples slowly after synthesis was expected to decrease the extent of disorder in the crystals and increase the average grain size.  In an attempt to characterized disorder, we turn again to the X-ray refinements.  There are at least four contributions to peak width in powder X-ray diffraction \cite{Cullity}: instrumental broadening, thermal vibrations, grain size, and microstrain. Instrumental broadening is a function of beam optics and geometry. Thermal vibrations increase the peak width with increasing temperature.  Peak width increases with reduced grain size and increasing microstrain. No variations in the X-ray peak widths are measured in the current pollycrystalline samples (inset of Fig. \ref{fig:SlowCoolxray}). Differential instrumental or thermal peak broadening can be ruled out, since all samples were prepared and measured at room temperature on the same instrument. Because all peaks are of similar width, no difference due to grain size or microstrain can be resolved between samples A, B and C.
 
\subsection{\label{sec:level2} Post synthesis annealing time $t$}

\begin{figure}[t!]
  \includegraphics[width=\linewidth]{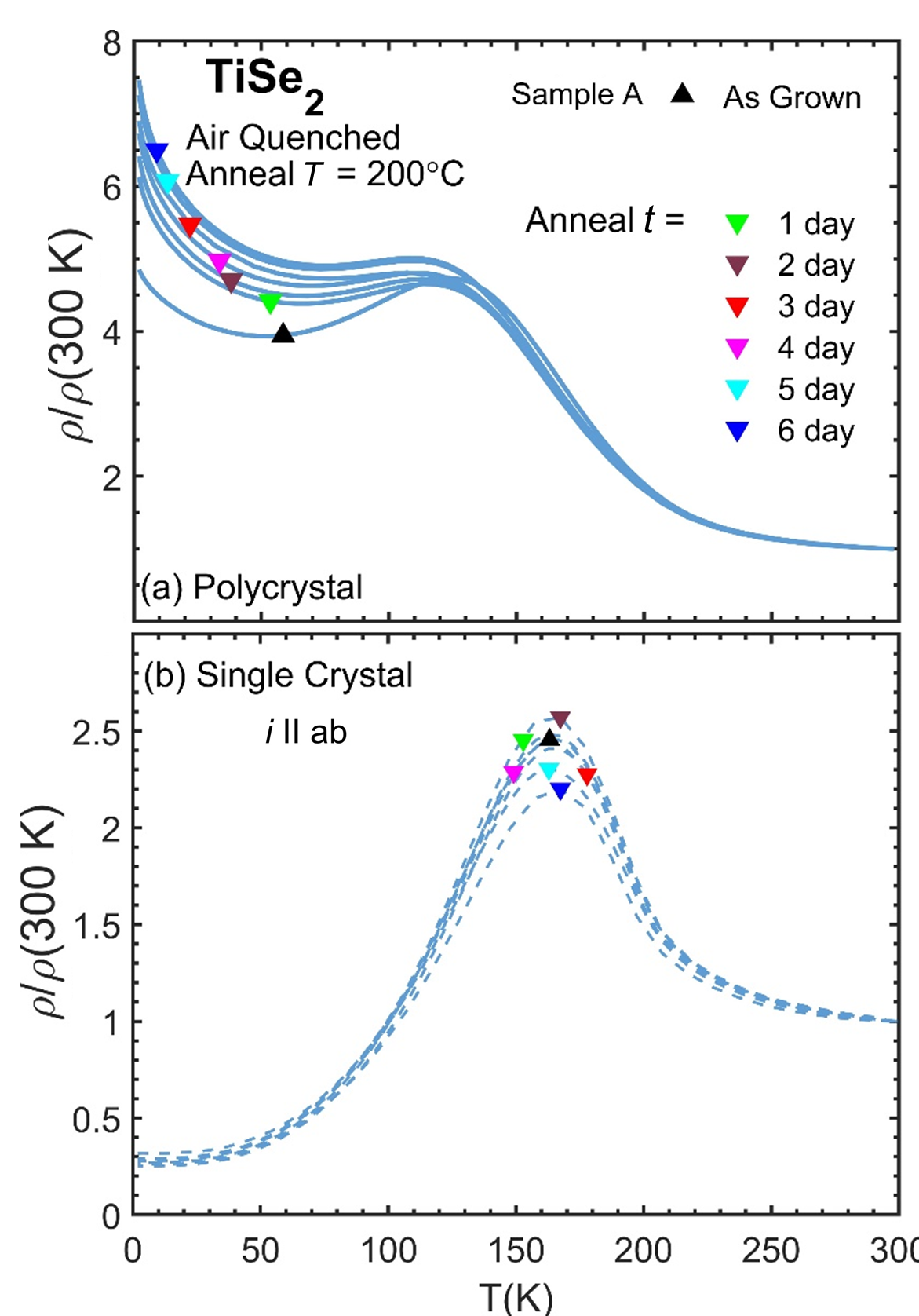}
  \caption{ (a) Comparison of polycrystalline $\rho/\rho(300$ K) for TiSe$_2$ samples  annealed at 200$^\circ$C  in one day increments up to six days.  With increase in anneal time, the low temperature resistivity increases. (b) The same comparison for TiSe$_2$ single crystals.  The low temperature resistivity is dominated by iodine impurities. }
  \label{fig:TimeStudy}
\end{figure}

The next set of experiments focuses on the effect of annealing time $t$. Different pieces of sample A were annealed at T = 200$^\circ$C, for times $t$ ranging from 1 to 6 days, followed by air quenching. The low anneal temperature was chosen to relieve quenched-in disorder without adding more disorder from quenching at a high temperature. Resistivity shows a general upward, albeit small trend at low temperatures for increasing $t$ (Fig. \ref{fig:TimeStudy}a). As before, no change is recorded in the X-ray peak width and lattice parameters (not shown). By comparison with the cooling rate \textit{r} (Fig. \ref{fig:SlowCoolResistivity} and Table \ref{Tab:EMPA1}), the change in the low temperature resistivity is much smaller when varying the annealing time \textit{t} at T = 200$^\circ$C: at the lowest measured temperature, the relative change in $\rho$ as a function of \textit{r} (Fig. \ref{fig:SlowCoolResistivity}) is $\rho_C/\rho_A~\sim$ 30, even for stoichiometry changes less than 1$\%$ (Table \ref{Tab:EMPA1}). The corresponding change in $\rho$ at low temperature with annealing time \textit{t} (Fig. \ref{fig:TimeStudy}) is $\rho(6days)/ \rho_A~\sim$ 1.5 with larger composition variation (Table \ref{Tab:EMPA3}). The latter reinforces the idea of the possibly intrinsic semiconductor state in TiSe$_2$, which is approached with longer annealing. Conversely, the role of stoichiometry variations, while unclear, appears to be minimal compared to the disorder and WL effects.

A similar study with anneal time was done on single crystals. The normalized $\rho$(T) is plotted in Fig. \ref{fig:TimeStudy}b.  Annealing did not change the low temperature transport properties when compared to the polycrystalline samples.  EPMA studies looking for only Ti and Se show all similar ratios as seen in Table \ref{Tab:EMPA3}.  However, as stated earlier, EPMA measurements reveal iodine inclusions around 1${\%}$ in single crystals for which the Ti:Se ratio is found to be 1:2. The additional density of states near the Fermi energy due to iodine accounts for the metal-like low temperature electrical transport down to 2 K in single crystalline TiSe$_2$.

\subsection{\label{sec:level2} Post synthesis annealing temperature $T$}

\begin{figure}[t!]
  \includegraphics[width=\linewidth]{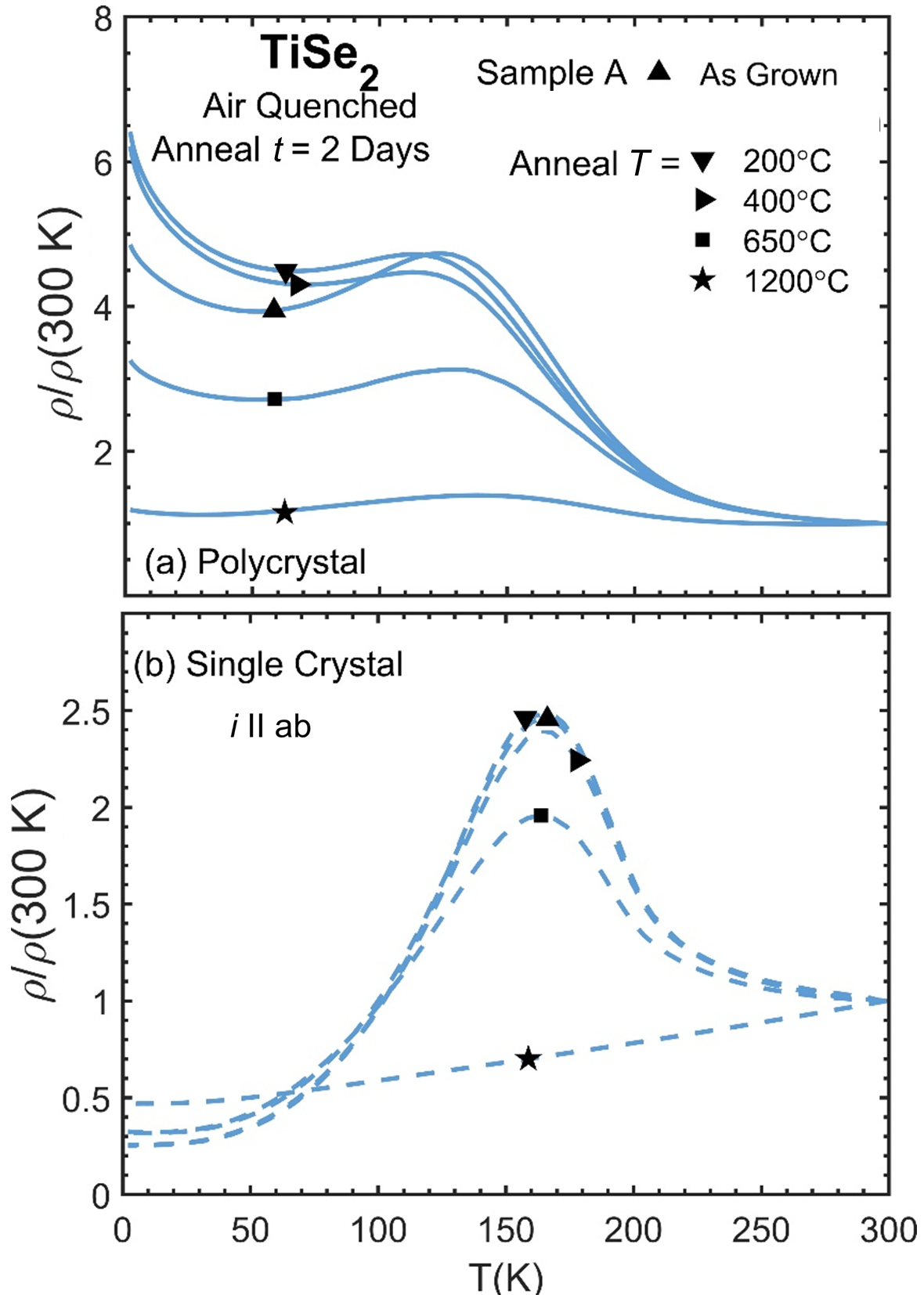}
  \caption{ (a) A comparison of normalized $\rho$(T) for polycrystalline TiSe$_2$ samples annealed at different temperatures post synthesis.   For anneal temperatures below the growth temperature, there is an increase in normalized $\rho$(T), while at higher temperature anneals, their is a decrease in low temperature normalized $\rho$(T). (b)  The corresponding study for single crystals.}
  \label{fig:TempDependenceResistivity}
\end{figure}

The next experiment aims to purposefully induce disorder into the TiSe$_2$ by quenching, followed by annealing at different temperatures $T$. 
Different single crystal pieces were annealed for 2 days at different temperatures $T$ between 200 and 1200$^\circ$C.  After annealing, all samples were quenched.  Normalized $\rho$(T) data is plotted in Fig. \ref{fig:TempDependenceResistivity}.

\begin{figure}
  \includegraphics[width=\linewidth]{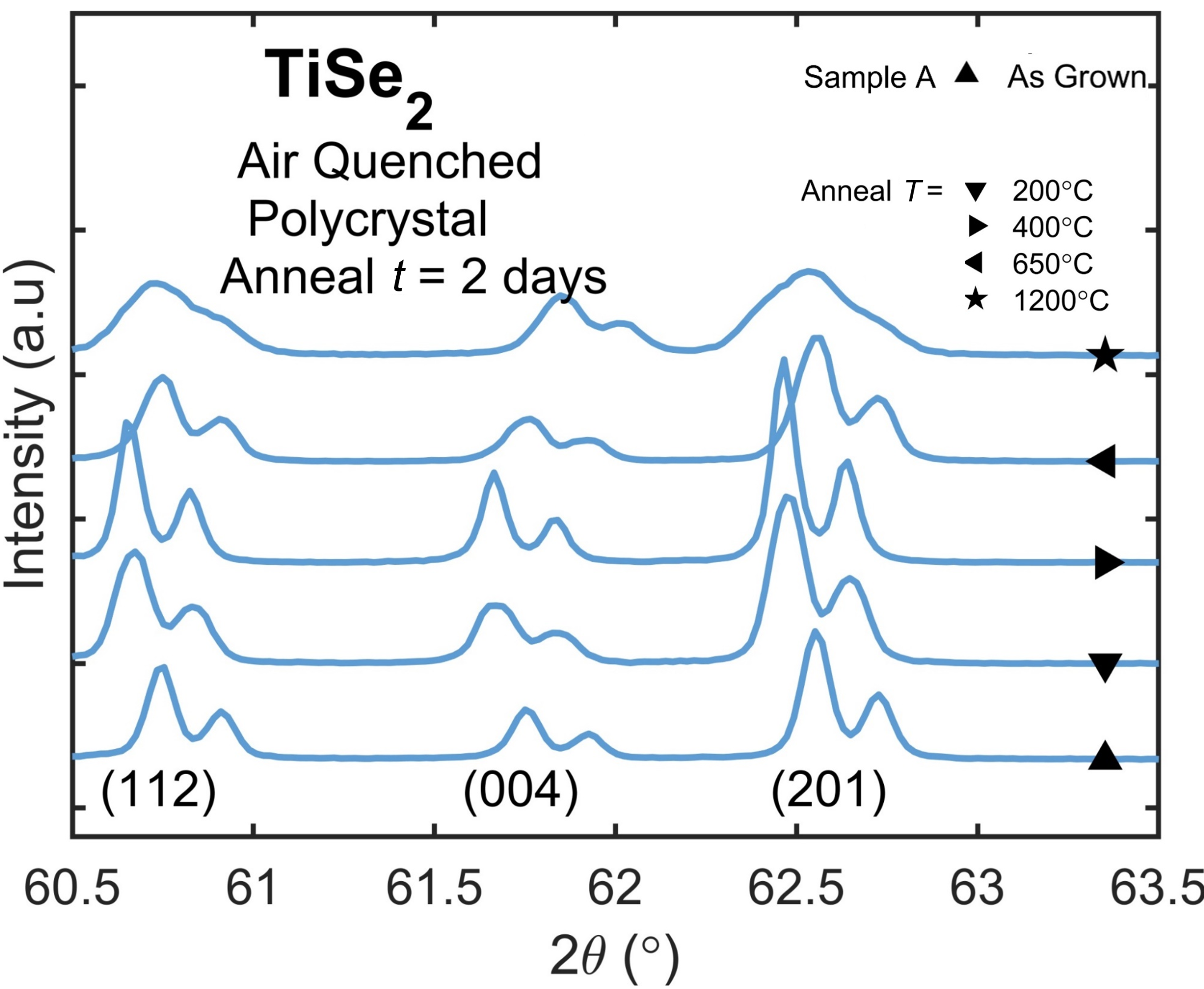}
  \caption{Zoomed in comparison of X-ray patterns for different anneal temperatures. At 1200$^\circ$C there is an increase in disorder from microstrain as evidence by severe broadening of the peaks.}
  \label{fig:xrayshift}
\end{figure}

For anneal temperatures $T$ below the growth temperature $T_{growth}$ = 650$^\circ$C (triangles, Fig. \ref{fig:TempDependenceResistivity}(a)), the low temperature resistivity of the polycrystalline samples increases compared to that of the as-grown sample, much the same as the result shown in Fig.~\ref{fig:TimeStudy}(a).  For anneal temperatures at (square) or above (star) the synthesis temperature, the low temperature resistivity decreases. However, below 20 K the resistivity increases on cooling for all annealing temperatures $T$. Our EPMA data shows no systematic loss of selenium with increased anneal temperature (Table \ref{Tab:EMPA4}), whereas X-ray diffraction patterns (Fig. \ref{fig:xrayshift}) indicate significant peak broadening for samples quenched from 1200$^\circ$C (star) indicating microstrain caused by quenching at such a high temperature. Though there are small variations in lattice parameters, the variations are less than 0.1$\%$ of the as grown (upwards triangle), so the change in resistivity is not due to a change in the unit cell.
	
For comparison, analogous data is shown in Fig. \ref{fig:TempDependenceResistivity}(b) for TiSe$_2$ single crystals. Besides the differences in low temperature resistivity, which can be explained by iodine impurities, the normalized $\rho$(T) shows qualitatively similar features as the polycrystalline samples. The trend of decreasing peak height below the CDW transition is qualitatively similar to that previously attributed to non-stoichiometry or disorder, or both\cite{Salvo1976,Hildebrand2016}. Though a Se deficiency is seen in the sample annealed at 1200$^\circ$C, the polycrystalline counterpart suggests that the decrease in the anomaly height is not due to doping, but rather an increase in quenched disorder. Remarkably, the 1200$^\circ$C single crystal (star, Fig. \ref{fig:TempDependenceResistivity}(b)) shows metallic behavior for the whole temperature range, and no anomaly in $\rho$(T). Consistent with the observations of the most substantive structural changes at this temperature (Fig. \ref{fig:xrayshift}), this signals that self doping, disorder, grain boundary freezing, or more, inhibit the plausible intrinsic semiconducting behavior of TiSe$_2$ at excessively high annealing temperatures.

\begin{center}
      \begin{table}
          \begin{tabular}{l || c c c }
              \multicolumn{3}{c}{} \\ \hline
               Anneal \textit{t} & Polycrystal  & Single Crystal  \\ 
              (days) & Se & Se  \\ \hline
              As Grown & ${2.00\pm0.02}$     & $2.013\pm 0.02$\\
              2 & ${2.03\pm0.02}$     & \\
              3 & ${2.03\pm0.03}$     &$2.017\pm 0.02$\\
              4 & ${2.01\pm0.02}$     & \\
              5 & ${1.98\pm0.02}$     &\\
              6 & ${2.00\pm0.03}$     & $2.012\pm 0.02$\\
              \hline \hline
          \end{tabular}
          \caption{Se normalized to 1 Ti in TiSe$_2$ samples annealed at 200$^\circ$C  in one day increments up to six days corresponding to Fig. \ref{fig:TimeStudy}.} \label{Tab:EMPA3}
      \end{table}
  \end{center}

	\begin{center}
      \begin{table}
          \begin{tabular}{l || c c c}
              \multicolumn{3}{c}{} \\ \hline
              Anneal \textit{T} ($^\circ$C) & Polycrystal & Single Crystal  \\ 
              {} & Se & Se \\ \hline
              As Grown          &${2.00\pm0.02}$    &${2.013\pm0.007}$\\
              200     &${2.03\pm0.05}$    &                \\
              400     &${2.02\pm0.04}$    &${2.04\pm0.01}$\\
              650    &${2.026\pm0.006}$  &${2.05\pm0.01}$\\
              1200    &${2.03\pm0.02}$    &${1.882\pm0.007}$\\
              \hline \hline
          \end{tabular}
          \caption{EPMA results for TiSe$_2$ samples annealed for 2 days at variable temperatures corresponding to Fig. \ref{fig:TempDependenceResistivity}.} \label{Tab:EMPA4}
      \end{table}
  \end{center}

In summary, our results on polycrystalline TiSe$_2$ are consistent with this system being a small band gap semiconductor at low temperatures. When synthesis conditions favor disorder, the  semiconducting behavior is concealed by enhanced metallicity. Though all polycrystalline samples in the current study are close to stoichiometry, the small band gap causes even the smallest deviations from the 1:2 stoichiometry to add impurity states, which, in turn, affect the low temperature transport. These impurity states become localized at low temperature, resulting in a logarithmic increase of the resistivity on cooling rather than the exponential increase expected from an activated gap. These observations are consistent with transport in polycrystalline TiSe$_2$ emerging from both semiconductor physics and localization physics, more commonly discussed in disorder metals.

\section{\label{sec:level1} Conclusions}

We have systematically studied the effects of the cooling rate \textit{r}, and temperature- and time-dependence \textit{T} and \textit{t} of post-synthesis annealing on the observed electrical transport properties of TiSe$_2$.  For the first time, the weak-localization effect is found in polycrystalline TiSe$_2$ samples, embodying the quantum corrections to the electrical transport properties in low dimensional systems.  At low temperatures results on polycrystalline TiSe$_2$ are consistent with a small gap semiconductor behavior, with low temperature $\rho(T)$ and MR dominated by the weak localization effect due to residual impurities. This study is intended to serve as a guide in the synthesis of TiSe$_2$, by pointing out the intrinsic and extrinsic properties as a function of the preparation method. 

\begin{acknowledgments}

\section{Acknowledgements}
JMM, JC, CLH and EM acknowledge support from  NSF DMREF grant 1629374. The use of the EPMA facility at the Department of Earth Science, Rice University, Houston, TX is kindly acknowledged. Furthermore, the authors are grateful for fruitful discussions with A. M. Hallas and M. D. Watson. 
\end{acknowledgments}
\bibliography{TiSe2lib}

\end{document}